\documentclass[twocolumn]{aastex631} 
\usepackage{CJK}
\usepackage{ulem}

\begin{document}
\begin{CJK*}{UTF8}{gbsn}
\title{The first quenched galaxies, when and how?}


\author[0000-0003-3864-068X]{Lizhi Xie (谢利智)}
\thanks{Email: xielizhi.1988@gmail.com}
\affiliation{Tianjin Normal University, Binshuixidao 393, Xiqing, 300387, Tianjin, China \\}
\affiliation{INAF – Astronomical Observatory of Trieste, via G.B. Tiepolo 11, I-34143 Trieste, Italy \\}

\author[0000-0002-6220-9104]{Gabriella De Lucia}
\affiliation{INAF – Astronomical Observatory of Trieste, via G.B. Tiepolo 11, I-34143 Trieste, Italy \\}
\affiliation{IFPU - Institute for Fundamental Physics of the Universe, via Beirut 2, 34151, Trieste, Italy\\}

\author[0000-0003-4744-0188]{Fabio Fontanot}
\affiliation{INAF – Astronomical Observatory of Trieste, via G.B. Tiepolo 11, I-34143 Trieste, Italy \\}
\affiliation{IFPU - Institute for Fundamental Physics of the Universe, via Beirut 2, 34151, Trieste, Italy\\}

\author[0000-0002-3301-3321]{Michaela Hirschmann}
\affiliation{Institute for Physics, Laboratory for Galaxy Evolution and Spectral Modelling, \\
Ecole Polytechnique Federale de Lausanne, Observatoire de Sauverny, Chemin Pegasi 51, 
CH-1290 Versoix, Switzerland}
\affiliation{INAF – Astronomical Observatory of Trieste, via G.B. Tiepolo 11, I-34143 Trieste, Italy \\}

\author[0000-0002-3196-5126]{Yannick M.~Bah\'{e}}
\affiliation{Laboratory for Astrophysics, 
\'{E}cole Polytechnique F\'{e}d\'{e}rale de Lausanne (EPFL), Observatoire de Sauverny, Chemin Pegasi 51, \\
CH-1290 Versoix, Switzerland}

\author[0000-0003-4849-9536]{Michael L. Balogh}
\affiliation{Department of Physics and Astronomy, University of Waterloo, Waterloo, Ontario N2L 3G1, Canada} 

\author[0000-0002-9330-9108]{Adam Muzzin}
\affiliation{Department of Physics and Astronomy, York University, 4700 Keele St., Toronto, Ontario MJ3 1P3, Canada}

\author[0000-0003-0980-1499]{Benedetta Vulcani}
\affiliation{INAF- Osservatorio Astronomico di Padova, Vicolo Osservatorio 5, 35122 Padova, Italy}

\author[0000-0002-8209-2783]{Devontae C. Baxter}
\affiliation{Department of Astronomy \& Astrophysics,
University of California, San Diego, 9500 Gilman Dr, La Jolla, CA 92093, USA \\}

\author[0000-0001-6003-0541]{Ben Forrest}
\affiliation{Department of Physics and Astronomy, University of California Davis, One Shields Avenue, Davis, CA 95616 USA}

\author[0000-0002-6572-7089]{Gillian Wilson}
\affiliation{Department of Physics, University of California Merced, 5200 Lake Road, Merced, CA 95353, USA}

\author[0000-0003-1371-6019]{Gregory H. Rudnick}
\affiliation{Department of Physics and Astronomy, University of Kansas, Lawrence, KS 66045, USA}

\author[0000-0001-5851-1856]{M. C. Cooper}
\affiliation{Department of Physics and Astronomy, University of California, Irvine, 4129 Frederick Reines Hall, Irvine, CA 92697, USA}

\author[0000-0002-9280-0173]{Umberto Rescigno}
\affiliation{Instituto de Astronom\'ia y Ciencias Planetarias de Atacama (INCT), Universidad de Atacama, Copayapu 485, Copiap\'o, Chile\\}
\affiliation{Instituto de Astrofisica, Universidad Andres Bello, Fernandez Concha 700, Las Condes, Santiago RM, Chile}

\begin{abstract}
Many quiescent galaxies discovered in the early Universe by \textit{JWST} raise fundamental questions on when and how these galaxies became and stayed quenched. Making use of the latest version of the semi-analytic model GAEA that provides good agreement with the observed quenched fractions up to $z\sim 3$, we make predictions for the expected fractions of quiescent galaxies up to $z\sim 7$ and analyze the main quenching mechanism. We find that in a simulated box of $685~{\rm Mpc}$ on a side, the first quenched massive ($M_{\star} \sim 10^{11} {\rm M}_{\odot}$), Milky Way mass, and low mass ($M_{\star} \sim 10^{9.5} {\rm M}_{\odot}$ ) galaxies appear at $z\sim 4.5$, $z\sim 6.2$, and before $z = 7$.
Most quenched galaxies identified at early redshifts remain quenched for more than 1 Gyr. 
Independently of galaxy stellar mass, the dominant quenching mechanism at high redshift is accretion disk feedback (quasar winds) from a central massive black hole, which is triggered by mergers in massive and MW-mass galaxies, and by disk instabilities in low-mass galaxies. Environmental stripping becomes increasingly more important at lower redshift. 
\end{abstract}


\section{Introduction} \label{sec:intro}

The cessation of star formation in galaxies has drawn considerable attention in recent years, especially given the large numbers of quiescent massive galaxies that have been found in the early Universe \citep{Schreiber2018, Merlin2019, Girelli2019, Glazebrook2017, Nanayakkara2024}, when the timescale available to assemble and quench these systems is short. 
Spectroscopic studies suggest that these galaxies experience short periods of intense star formation, grow up to a stellar mass of $10^{11} {\rm M}_{\odot}$ in the first one or two billion years of the universe, and then stop forming stars within a few tens of million years
\citep{Forrest2020, Valentino2020, Kakimoto2023, Carnall2023b}.
This rapid assembly and quenching process might challenge our current understanding of galaxy formation \citep{Finkelstein2023}.

The number densities of quenched massive galaxies $M_{\star} > 10^{10.5} {\rm M}_{\odot}$ increase rapidly from  $\sim 10^{-6}~{\rm Mpc}^{-3}$ at $z\sim 5$ to as much as a factor of $10$ times higher at $z\sim 3$ \citep{Marsan2022}, although the measured number densities have a relatively large scatter due to different selection criteria and cosmic variance \citep{Valentino2023}. The classical UVJ color selection of quenched galaxies  \citep{wuyts2008} is found to be incomplete and underestimates the number of quenched galaxies at $z>3$ \citep{Schreiber2018}. Some studies propose a modified UVJ selection \citep{Belli2019}, others favour a NUVrJ  \citep{Ilbert2013} or ugi color selection \citep{Antwi-Danso2023}, to identify galaxies that have been quenched recently, which is important for galaxies at $z>3$ \citep{Gould2023, Kubo2024}. 

Despite a large scatter in observational measurements, it is a solid conclusion that most theoretical models under-predict the number densities of quenched galaxies \citep{Cecchi2019, Girelli2019, Gould2023} at $z>4$ by about an order of magnitude.   
\citet{Weaver2023} found that quenched galaxies in the SHARK model \citep{Lagos2023} and IllustrisTNG simulation \citep{Pillepich2018} at $3.5<z<4.5$ are not as massive as the observed ones. 
Either creating enough massive galaxies at early cosmic epochs or quenching them on a short time scale remains a challenge for current galaxy formation models.

Various physical mechanisms have been proposed to explain the rapid assembly of mass in the early Universe, including weaker feedback \citep{Dekel2023}, enhanced star formation efficiencies \citep{Wang2023}, and a top-heavy IMF \citep{Trinca2023}. The physical mechanisms driving quenching also remain unclear. Quenching could be caused by internal feedback from active galactic nuclei (AGN) and supernovae (SN) feedback, or by external physical processes including galaxy-galaxy interactions and environmental stripping. With a minimal physical model, \citet{Gelli2023} suggests that SN feedback is not powerful enough to quench galaxies of $\sim 10^{8} {\rm M}_{\odot}$ at high redshift. 
The fact that many high-$z$ quenched galaxies are found to host luminous AGN \citep{Ito2022, Shimakawa2023, Carnall2023b, Belli2023, Davies2023, DEugenio2023} suggests an important contribution to quenching from feedback from their central supermassive black holes (SMBH). This appears to be confirmed in recent theoretical works: 
\cite{Kurinchi-Vendhan2023} and \cite{Kimmig2023} analyze the massive quenched galaxies in TNG and Magneticum at $z\sim 3$ and show that these galaxies are indeed quenched by AGN feedback. \citet{Lovell2023} found that AGN feedback is the dominant quenching mechanism for galaxies above $10^9 {\rm M}_{\odot}$ at $z\sim 5$. \citet{Qin2017} use semi-analytic models to identify analogues of quenched galaxies observed at $z\sim 5$ and show that these have grown through significant mergers and host the most massive black holes at their redshifts. 
Some quenched galaxies are found in over-dense environments \citep{Kubo2021, McConachie2022, Tanaka2023, Alberts2023}, suggesting environmental quenching may also contribute as early as $z\sim 4$.

In our recent work \citep[][here after GAEA2023]{DeLucia2024}, we present the latest version of our GAlaxy Evolution and Assembly (GAEA) model and show that it can correctly reproduce the observed quenched fractions up to redshift $\sim 3$ as well as the number densities of quenched galaxies up to redshift $\sim 5$, better than many state-of-the-art models and simulations. The good agreement with observations makes it a perfect tool for studying the physical origin of quenched galaxies at high redshift.  

In this work, we extend the analysis presented in \citet{DeLucia2024} to the quenched fractions and their quenching mechanisms since $z\sim 7$. In Section~\ref{sec:model}, we introduce the semi-analytic model and our methodology. In Section~\ref{sec:results} and Section~\ref{sec:conlusion}, we present the results and give our conclusions.

\section{Model and Simulation} \label{sec:model}

GAEA2023 \citep{DeLucia2014} now combines independent versions of the model including an improved treatment for the supernovae feedback \citep{Hirschmann2016}, of the multi-phase cold gas \citep{Xie2017}, of environmental effects \citep{Xie2020}, and of AGN accretion and feedback \citep{Fontanot2020}. In particular, GAEA2023 implements a treatment for tidal stripping and ram pressure stripping that gradually removes hot gas, as well as ram-pressure stripping of cold gas, for satellite galaxies. These implementations help to solve the over-quenching of low-mass satellite galaxies and to improve the model predictions in terms of gas fractions \citep{Xie2020}. GAEA2023 also implements updated modeling for cold gas accretion onto supermassive black holes. Mergers and disk instability cause a fraction of cold gas to lose angular momentum and flow towards the center, where it forms an accretion disk around the supermassive black hole. This material is then accreted onto the SMBH on a viscous timescale: accretion can heat the surrounding gas and cause an outflow (for details about the model, we refer to \citealt{Fontanot2020}). Below, we refer to this process as accretion disk feedback. GAEA2023 has been calibrated to reproduce the galaxy stellar mass function up to $z\sim 3$, HI mass function and quenched fractions at $z\sim 0$, as well as AGN luminosity function up to $z\sim 4$. In \citet{DeLucia2024} we show that this model version reproduces well the observed quenched fraction, the stellar mass function of the quenched population up to $z\sim 3$, as well as the number densities of quenched massive galaxies at up to $z\sim 5$.

The model is run on the Millennium Simulation \citep{Springel2005} with a box size of $685$ Mpc based on a WMAP 1-yr cosmology \citep{Spergel2003} with $\Omega_m = 0.25$, $\sigma_b = 0.045$, $\sigma_8 = 0.9$, and $h = 0.73$. 

In the following, we will compare GAEA2023 results with predictions from TNG100 and TNG300 \citep{Springel2018, Nelson2018, Naiman2018, Marinacci2018, Pillepich2018}. The TNG project is a suite of cosmological magneto-hydro-dynamical simulations, adopting the Planck cosmology \citep{Planck2016} with $\Omega_m = 0.3089$, $\Omega_b = 0.0486$, $\sigma_8 = 0.8159$, and $h = 0.6774$. The TNG100 and TNG300 simulate cubic boxes of side lengths approximately 100 and 300 Mpc. TNG considers two modes of AGN feedback: for high accretion rates, the surrounding gas is heated by thermal feedback from AGN. When the accretion rates are low, gas instead receives a kinetic `kick' that causes gas outflows. In this work, we use the publicly available database\footnote{https://www.tng-project.org/data/} to retrieve the simulated stellar mass, instantaneous star formation rate, and black hole mass within twice the stellar half-mass radius.

\section{Results}
\label{sec:results}

\subsection{Quenched fraction} \label{subsec:fq}

\begin{figure*}[ht!]
\centering
\includegraphics[width=0.8\textwidth]{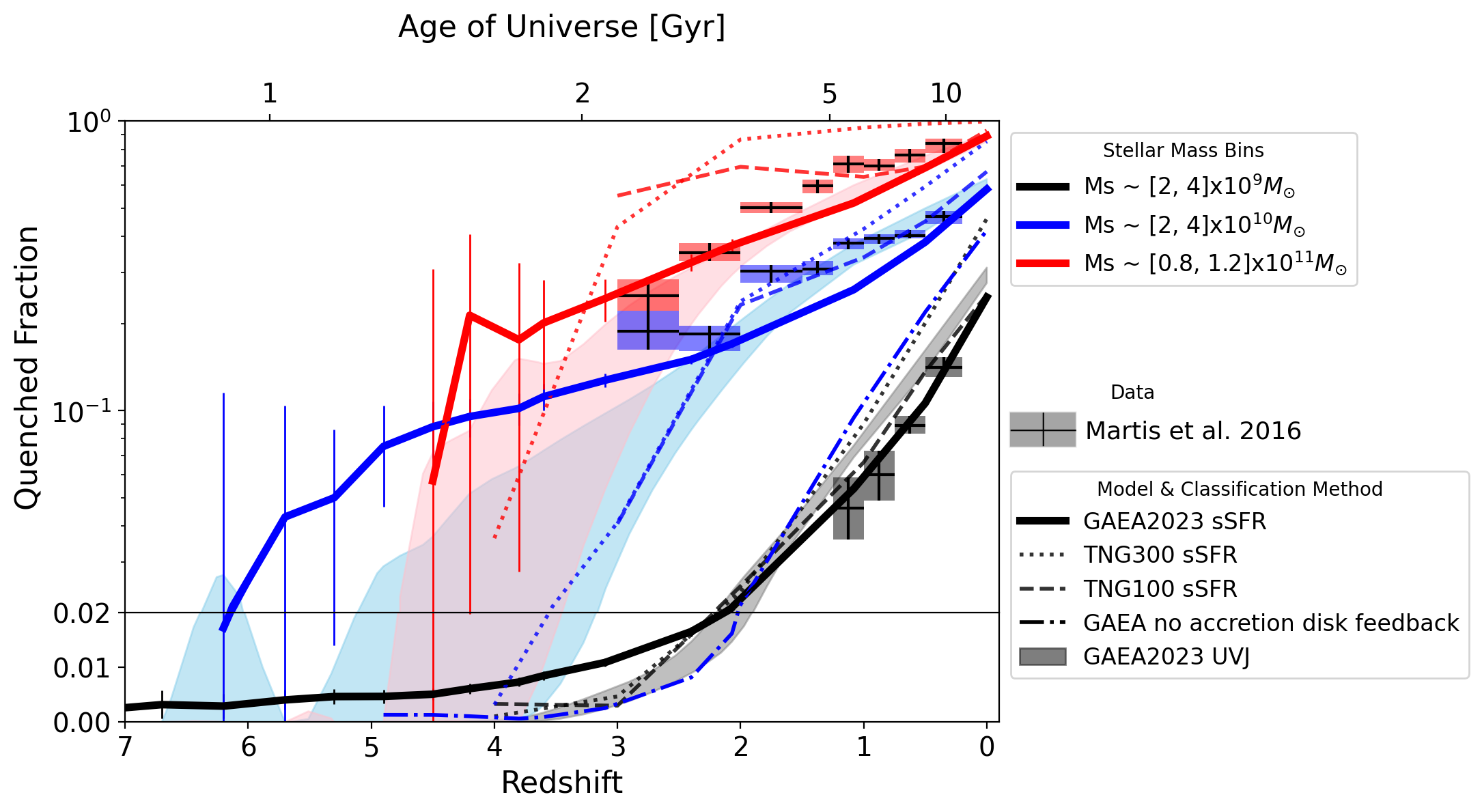}
\caption{Quenched fractions as a function of redshift. Different colors represent galaxies of different stellar masses. Vertical error bars show the standard deviations obtained for 100 randomly selected sub-samples. Shaded regions show the uncertainties in quiescent fractions from slightly different cuts in the UVJ diagram (more details in text). The dash-dotted line shows results from a GAEA run where accretion disk feedback is switched off. Dotted and dashed lines show the quenched fractions measured from TNG300 and TNG100. Error boxes are observed quenched fractions for UltraVISTA DR1 and 3D-HST survey from \citet{Martis2016}.
\label{fig:fq}}
\end{figure*}

Figure~\ref{fig:fq} shows the evolution of quenched fractions as predicted by GAEA2023 for low-mass ($2-4\times 10^{9} {\rm M}_{\odot}$), Milky-Way-mass (MW-mass, $2-4\times 10^{10} {\rm M}_{\odot}$), and massive galaxies ($0.8-1.5\times 10^{11} {\rm M}_{\odot}$). We consider different definitions for quenched galaxies: first, we select a sample imposing specific star formation rate $sSFR = SFR/M_{\star} < 0.3/t_H$ \footnote{We have verified that using a flat cut $sSFR < 10^{-11}~{\rm yr}^{-1}$, our predicted quenched fractions do not vary significantly \citep[see also in ][]{DeLucia2024}.}, where $t_H$ is the Hubble time at given redshift \citep{Franx2008}. 
It is important to stress that the star formation rates from GAEA2023 are averaged over the time interval between two snapshots, which is $\sim 80$ Myr at $z\sim 7$ and increases to $\sim 300$ Myr at $z\sim 0$. Therefore, our sample of model quenched galaxies does not include those that are only temporarily quenched by e.g. bursty periods of star formation. 
To get a fair comparison with observational data, we also present quenched fractions based on a UVJ (synthetic) color selection. The magnitudes are computed by convolving the star formation and chemical evolution history with photometric tables from \citet{bc03} and accounting for dust attenuation \citep{DeLucia2007}. We tried different cuts commonly adopted in the literature \citep{Williams2009, Whitaker2011, Muzzin2013, Martis2016} and plot the scatter obtained as shaded regions in Fig.~\ref{fig:fq}. For massive and MW-mass galaxies at $z>1$, we also used the diagonal selection cut proposed by \citet{Belli2019}, which is designed to select recently quenched galaxies. 
The error boxes show observational measurements of quenched fractions for galaxies from the UltraVISTA DR1 and 3D-HST surveys \citep{Martis2016}. In the redshift range $0<z<3$, there is a good agreement between GAEA2023 and data at all mass scales. The quenched fraction defined using the synthetic UVJ photometry are similar to those defined by sSFR at $z<2$, but decrease more rapidly at $z>2$. As for GAEA2023, the UVJ color selection underestimates the fraction of quiescent galaxies at $z>2$.
For TNG the quenched fractions of galaxies with $M_{\star} \sim 10^{11} {\rm M}_{\odot}$ are larger than observational measurement. 

Moving to higher redshifts, the predicted quenched fractions decrease. In the framework of GAEA2023, the quenched fraction of low-mass galaxies is 0.2 per cent at $z\sim 7.3$, \textit{i.e.} 6 out of 2880 galaxies are quenched. The quenched fraction remains below 1 per cent until $z\sim 3$. We traced 42 quenched low-mass galaxies at $z\sim 6.2$ forward in time and found that 11 of them moved back to the main sequence within 0.5 Gyr. Most of the high-redshift quenched low-mass galaxies, however, remain quiescent for more than 1 Gyr. 

The first quenched galaxies with a mass similar to the Milky Way appear at $z\sim 6.2$, \textit{i.e.} 2 out of 116 MW-mass galaxies are quenched. One of these two returns to the main sequence after $\sim 0.5$ Gyr. The other is a satellite that remains quenched until it merges with another galaxy. The quenched fraction, for galaxies of this mass, grows quickly to $10$ per cent between $4<z<5$. 

The first massive quenched galaxies (with $M_{\star} \sim 10^{11}{\rm M}_{\odot}$) are found at $z\sim 4.5$. 2 out of 35 galaxies in this mass bin are quenched at this redshift. In their subsequent evolution, these two galaxies continue to have a low star formation rate for most of the time.

Predictions from TNG are quite different from those based on GAEA2023, with systematically lower quenched fractions at high redshift, and no quenched galaxy at $z>4$. This is in clear tension with the existence of spectroscopically confirmed quenched massive galaxies at $z>4$\citep[e.g.][]{Carnall2023a}. A similar result is reported in \citet{Merlin2019}, where a lower sSFR cut $10^{-11}~{\rm yr}^{-1}$ was used. Though most massive simulated galaxies at $z\sim 4$ have a central massive black hole of $10^{8} {\rm M}_{\odot}$, only a minor fraction of them are quenched. In TNG, kinetic feedback from AGN represents the most efficient quenching mechanism for massive galaxies  \citep{Kurinchi-Vendhan2023}. However, the accretion rates at high redshift are so large that the assumed mode for AGN feedback in TNG comes in the form of thermal feedback, which is not efficient enough to quench galaxies at $z>3$.

\subsection{Quenching Mechanisms} \label{subsec: why}

\begin{figure}    
    \includegraphics[width=0.45\textwidth]{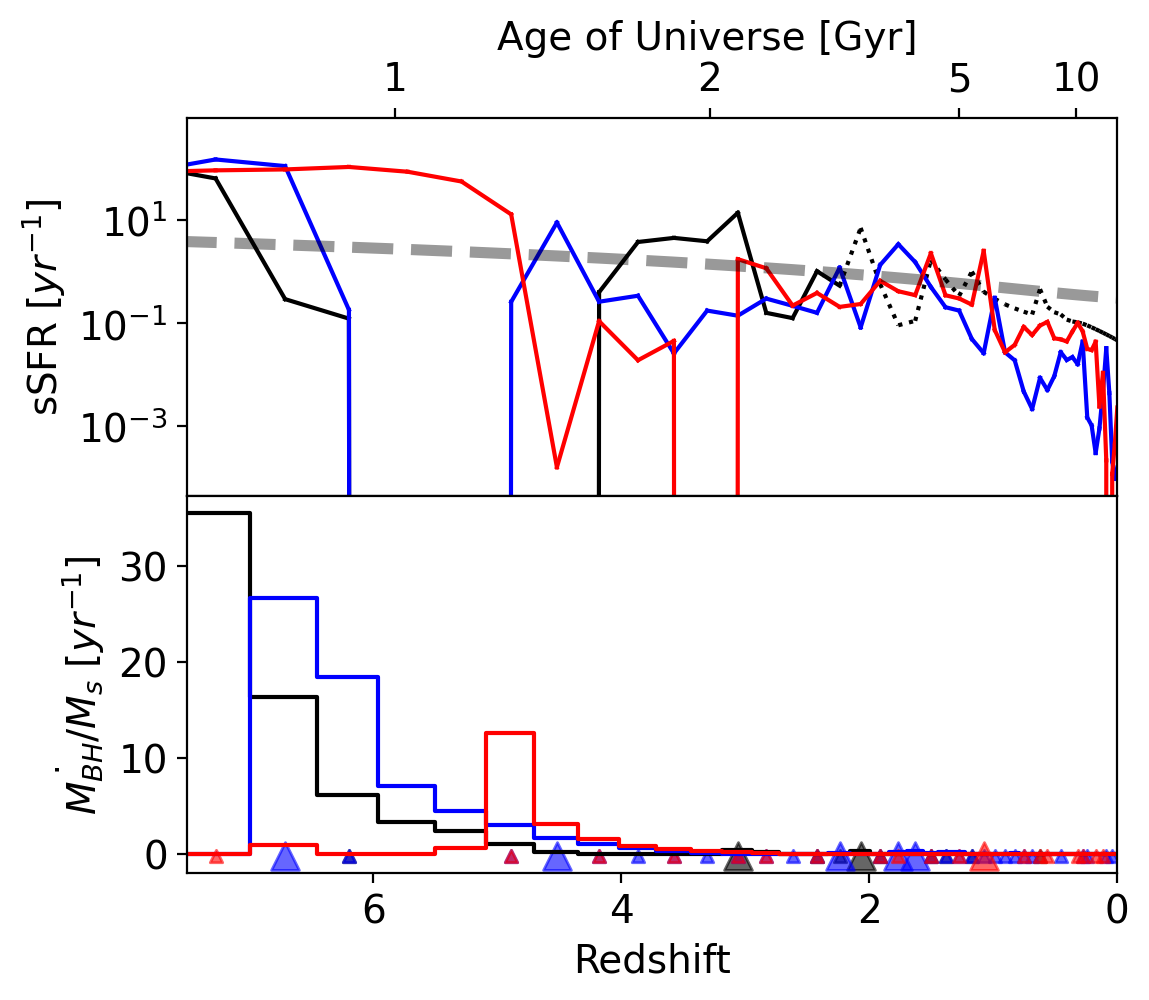}
    \caption{Evolution of three representative galaxies. The upper and lower panels show the evolution of sSFR and the SMBH accretion rate associated with the disk accretion mode normalized by stellar mass. The sSFR is indicated by solid and dotted lines when a galaxy is classified as central or satellite, respectively. The gray straight line in the upper panel is the separation between quenched and star-forming galaxies. Large and small triangles in the lower panel indicate merger events with mass ratios above 0.3 and 0.01. Color code is the same as in Figure~\ref{fig:fq}.}
    \label{fig:Evo}
\end{figure}

Broadly speaking, we can consider two quenching scenarios:  internal quenching, i.e. AGN feedback and SN feedback, and external quenching, i.e. environmental stripping and galaxy-galaxy interactions. In this section, we analyze the relative importance of these quenching mechanisms at different redshifts in the GAEA framework.

First of all, we traced the history of high-redshift quenched galaxies in the three stellar mass ranges considered. Figure~\ref{fig:Evo} shows the evolution histories of three representative galaxies selected from each stellar mass bin.
All quenched model galaxies have experienced high-rate black hole accretion and suffered accretion disk feedback right before quenching, suggesting that this mechanism is the main quenching channel at this redshift. This is confirmed by the dot-dashed lines in Figure~\ref{fig:fq}, showing predictions for MW-mass galaxies from the version of GAEA \citep{Xie2020} that does not include disk accretion feedback and that significantly under-predicts the fractions of quenched galaxies at high redshift \citep[see also in][]{DeLucia2024}.

\begin{figure}[ht!]
\includegraphics[width=0.45\textwidth]{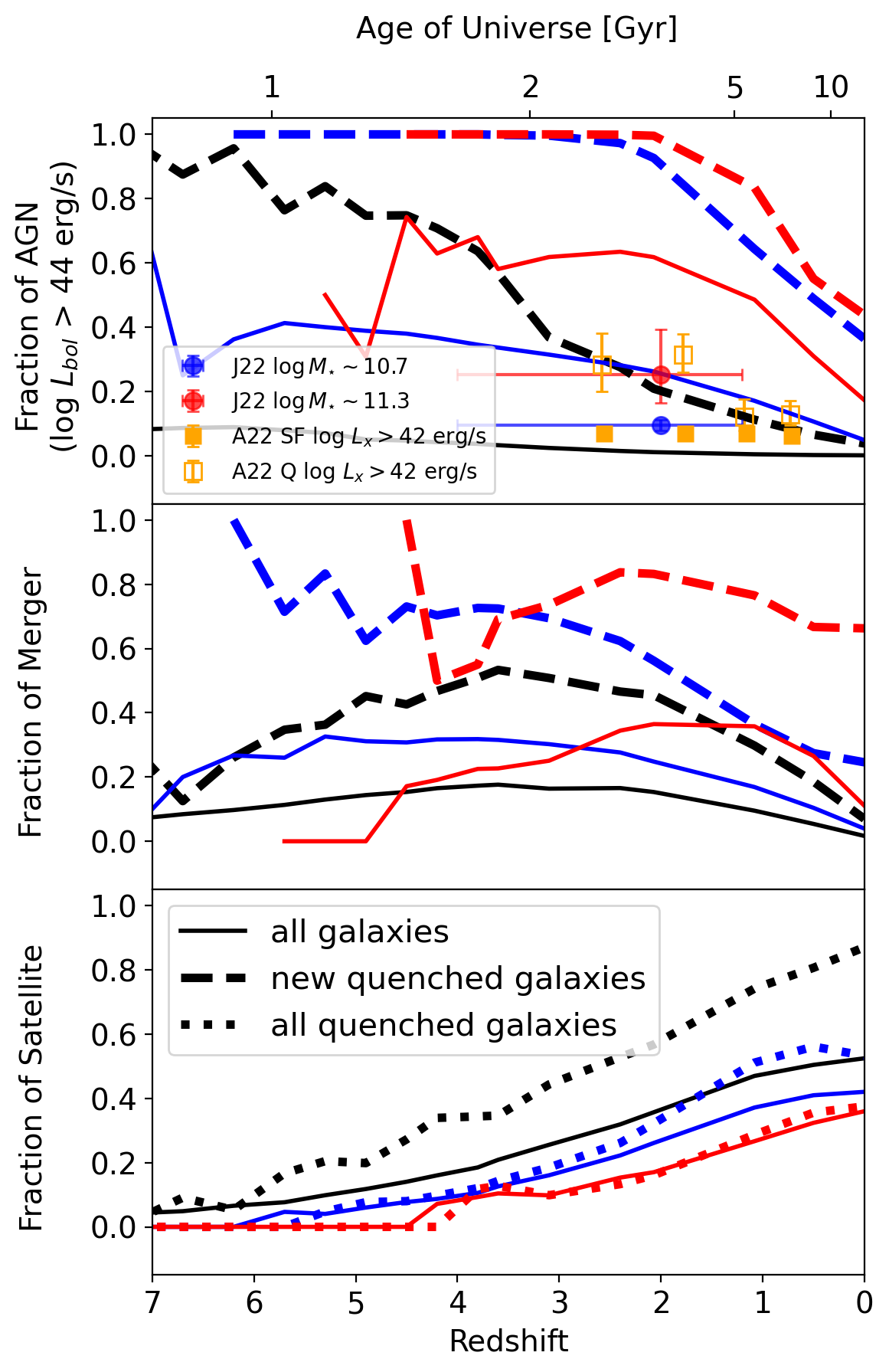}

\caption{The top, middle, and bottom panels show the fraction of AGN hosts ($\log L_{bol}/[{\rm erg/s}]>44$), the fraction of galaxies that have experienced recent mergers with a mass ratio larger than 1/100, and the fraction of satellites, respectively. Solid, dotted, and dashed lines correspond to the total, quenched, and newly-quenched galaxy populations. Color code is the same as in Figure~\ref{fig:fq}. 
Red and blue circles show observationally estimated AGN fractions with $44 < \log L_{bol}/[{\rm erg/s}] < 45.5$ for massive and MW-mass galaxies \citep{Ji2022}. Orange open and filled squares show AGN fractions with $\log L_{X}/[{\rm erg/s}] >42$ for star-forming and quenched galaxies \citep{Aird2022}. These studies are based on X-ray-selected AGN fractions and should be intended as lower limits for the overall AGN fractions. The model predictions, however, are the upper limits of the actual AGN population.
\label{fig:fbh}}
\end{figure}

Since large accretion rates give rise to luminous quasars, in the top panel of Figure~\ref{fig:fbh} we compare the fraction of quasar-host galaxies in recently-quenched galaxies (dashed) and the entire population (solid). 
The `recently quenched’ galaxies are those that got quenched since the last snapshots, namely quenched in the previous $\sim 80$ Myr at $z\sim7$ and $\sim 300$ Myr at $z=0$. The BH accretion rate that is used to calculate the bolometric luminosity is also averaged on the same timescale.

About 60 and 30 per cent of massive and MW-mass galaxies host AGN with bolometric luminosity brighter than $10^{44}$ erg/s at $z>2$. These fractions rise to 100 per cent for recently quenched galaxies at $z>2$.  
Surprisingly, more than $70\%$ of low-mass recently quenched galaxies at $z>4$ also host luminous AGN, whereas the fraction is only $10\%$ for all galaxies in this mass range. The elevated fraction of luminous AGN for quenched galaxies confirms that the disk accretion feedback from SMBH is the dominant quenching mechanism for high redshift galaxies in the framework of GAEA2023. 
The fraction of luminous AGN decreases at lower redshift, and the differences between all galaxies and quenched samples also reduce: at low redshift, accretion disk feedback is less important for quenching. 

When comparing to observations, the model predicted AGN fractions are larger than the observational measurements for X-ray-selected AGN at $z<3$ \citep{Ji2022, Aird2022}. 
However, this is not surprising given that the model predicted AGN fractions represent upper limits to the actual AGN population, as many of these AGN events may have expired within the timescale we use to estimate the average BH accretion rate. This is especially true for low redshift, where the timescale is much longer and leads to an overestimate of AGN hosts concerning observational measurements. Additionally, the observational measurements should be intended as lower limits and can be significantly affected by selection, obscuration \citep{Hickox2018}. 
We also note that there have been many recent studies reporting discoveries of quenched galaxies hosting AGN\citep{Ito2022, Carnall2023b, DEugenio2023, Davies2023}, as well as several cases where AGN emission is absent\citep{Nanayakkara2024, Jin2023}, leaving this issue under debate.

The disk accretion feedback is triggered by both disk instabilities and mergers. While tracing the evolution of individual galaxies, we find that large accretion rates are associated with merger events for MW-mass and massive galaxies. In the middle panel of Figure~\ref{fig:fbh} we show the fraction of galaxies that just experienced mergers between the recently quenched and the entire population. 
All mergers with a mass ratio larger than $1:100$ are considered, motivated by the rapid growth of black holes driven by multiple minor mergers or even very minor mergers, which we find to be common for high-z quiescent galaxies. 
Compared to the entire sample of MW-mass and massive galaxies, newly quenched galaxies have much higher merger rates, especially at high redshift. Therefore, mergers represent the main channel for black hole accretion at these mass ranges in the GAEA framework.

For low-mass galaxies, more than half of the recently quenched galaxies haven't experienced any mergers around the time of quenching. We thus conclude that their SMBH accretion is not primarily driven by mergers. We find that low-mass galaxies are more likely to have unstable disks at high redshift, and it is this disk instability that triggers efficient black hole accretion. 
 
The connection between the quenching process and black hole accretion becomes weaker at lower redshift, where environmental effects become increasingly important. The bottom panel of Figure~\ref{fig:fbh} shows the fraction of galaxies that are satellites. In GAEA2023, satellite galaxies lose hot gas and cold gas gradually by tidal stripping and ram-pressure stripping, whereas central galaxies are unaffected. For low-mass galaxies, a larger fraction of satellite galaxies are quenched at as early as $z\sim 6$, so the dependence of quenching on the environment starts 
at very early epochs in the framework of our model. The phenomenon of environmental quenching starts since $z\sim 3$ for Milky-Way-mass galaxies. The difference in satellite fraction between quenched and all massive galaxies is negligible, which is consistent with previous results that massive galaxies are mainly quenched by AGN feedback \citep{Xie2020, Kimmig2023, Lovell2023, Qin2017}.

\section{Conclusion} \label{sec:conlusion}

In this work we use the semi-analytic model GAEA2023 to study the quenched fractions predicted for massive ($M_{\star} \sim 10^{11} {\rm M}_{\odot}$), MW-mass ($M_{\star} \sim 10^{10.5} {\rm M}_{\odot}$), and low mass galaxies ($M_{\star} \sim 10^{9.5} {\rm M}_{\odot}$) since $z= 7$. GAEA2023 predictions are in good agreement with the observed quenched fraction measured from UltraVista and 3D-HST in the redshift range $0<z<3$. 

The quenched fractions defined by UVJ color are consistent with those defined by sSFR up to $z\sim 3$. At higher redshift, the quenched fractions are under-estimated by a UVJ color selection. When adopting a sSFR selection, about $5 \%$ of massive galaxies are found to be quenched at $z\sim 4.5$, about $2\%$ of MW-mass galaxies are firstly found to be quenched at $z\sim 6.2$, and the quenched fraction of low-mass galaxies is $0.2\%$ at $z\sim 7$. More than half of galaxies maintain a low star formation rate for over $\sim 1$ Gyr.

All recently quenched MW-mass and massive galaxies at $z>2$, and more than 60 per cent of low mass newly quenched galaxies at $z>4$ host luminous quasars (with bolometric luminosity brighter than $10^{44}$ erg/s). This suggests that accretion disk feedback from SMBHs is the main reason for quenching at high redshift, as confirmed by analyzing predictions from an alternative model where this physical process is switched off. We find that disk accretion feedback responsible for quenching is driven by galaxy mergers for massive and MW-mass galaxies, and by disk instabilities for lower mass galaxies. Environmental effects become increasingly important for low-mass galaxies at $z<6$, and for MW-mass galaxies at $z<2$. Massive galaxies are not quenched by environmental processes in the framework of GAEA\citep{Hirschmann2016, DeLucia2019}.

The earliest quenched galaxy so far is at $z \sim 7$ \citep{Looser2023}, which has a stellar mass of $M_{\star} \sim 5\times 10^{8} {\rm M}_{\odot}$. At $z\sim 5$, most quenched galaxies discovered are massive galaxies. Based on our model predictions, we expect to find non-negligible numbers of quenched galaxies with stellar mass $\sim 10^{9.5} {\rm M}_{\odot}$ at $z\sim 7$ or even higher redshift. A large fraction of these galaxies are expected to host luminous quasars.

\section*{Acknowledgments}
We thank the anonymous referee for insightful comments and suggestions that enabled us to improve the manuscript greatly.
LZX acknowledges support from the National Natural Science Foundation of China (No. 12041302, No. 11903023). 
This research was supported by the International Space Science Institute (ISSI) in Bern, through ISSI International Team project 543 "Understanding the evolution and transitioning of distant proto-clusters into clusters".  We are grateful for the support of ISSI and the use of their facilities. We also gratefully acknowledge the Lorentz Center in Leiden (NL) for facilitating discussions on this project.  BV acknowledges the support from the INAF Mini Grant 2022 `Tracing filaments through cosmic time' (PI Vulcani).  MLB acknowledges support from an NSERC Discovery Grant. GHR acknowledges support from NSF Astronomy and Astrophysics grant 2206473, HST grant GO-16300.004-A, and NASA ADAP grant 80NSSC19K0592. GW gratefully acknowledges support from the National Science Foundation through grant AST-2205189 and from HST program number GO-16300. YMB gratefully acknowledges funding from the Netherlands Organization for Scientific Research (NWO) under Veni grant number 639.041.751 and financial support from the Swiss National Science Foundation (SNSF) under funding reference 200021\_213076.



\bibliography{ref.bib}
\bibliographystyle{aasjournal}

\end{CJK*}
\end{document}